# Anisotropic Proton and Oxygen Ion Conductivity in Epitaxial $Ba_2In_2O_5$ Thin Films

*Aline Fluri [a], Maths Karlsson [b], Marco Bettinelli [c], Ivano E. Castelli [d], Thomas Lippert [a,e]\*, Daniele Pergolesi [a]\**

[a] Thin Films and Interfaces Group, Research with Neutrons and Muons Division, Paul Scherrer Institute, 5232 Villigen PSI, Switzerland

[b] Department of Physics, Chalmers University of Technology, 412 96 Gothenburg, Sweden.

[c] Luminescent Materials Laboratory, University of Verona, 37134 Verona, Italy.

[d] Department of Chemistry, University of Copenhagen, DK - 2100 Copenhagen Ø, Denmark.

[e] Department of Chemistry and Applied Biosciences, Laboratory of Inorganic Chemistry, ETH Zürich, CH-8093 Zürich, Switzerland

\* Corresponding authors: daniele.pergolesi@psi.ch, thomas.lippert@psi.ch

**Abstract**

Solid oxide oxygen ion and proton conductors are a highly important class of materials for renewable energy conversion devices like solid oxide fuel cells. $Ba_2In_2O_5$ (BIO) exhibits both oxygen ion and proton conduction, in dry and humid environment, respectively. In dry environment, the brownmillerite crystal structure of BIO exhibits an ordered oxygen ion sublattice, which has been speculated to result in anisotropic oxygen ion conduction. The hydrated structure of BIO, however, resembles a perovskite and the protons in it were predicted to be ordered in layers.

To complement the significant theoretical and experimental efforts recently reported on the potentially anisotropic conductive properties in BIO, we measure here the proton and oxygen ion conductivity along different crystallographic directions.

Using epitaxial thin films with different crystallographic orientations the charge transport for both charge carriers is shown to be anisotropic. The anisotropy of the oxygen ion conduction can indeed be explained through the layered structure of the oxygen sublattice in brownmillerite BIO. The anisotropic proton conduction however, further supports the suggested ordering of the protonic defects in the material. The differences in proton conduction along different crystallographic directions attributed to proton ordering in BIO are of a similar extent as those observed along different crystallographic directions in materials where proton ordering is not present but where protons find preferential conduction pathways through chain-like or layered structures.

**Introduction**

Solid state oxygen ion and proton conducting thin films are currently of high interest, because of their applicability as electrolyte and electrode materials in present-day and future environmentally friendly solid oxide fuel cell technologies[1-2]. Some of these materials show a (defective) perovskite-related crystal structure, for example layered perovskite cathodes[3] or acceptor-doped proton conducting electrolytes[4], where oxygen vacancies are needed either to allow oxygen ion migration or proton uptake from a humid atmosphere. $Ba_2In_2O_5$ (BIO) shows predominant oxygen ion or proton conductivity, depending on the gaseous environment and temperature range[5-9]. The structure of BIO is related to a perovskite, but due to the charge of $Ba^{2+}$ and $In^{3+}$ cations, not all oxygen sites of a perovskite can be occupied. Up to around 900°C, BIO exhibits the brownmillerite structure (a=6.0864 Å, b=16.7903 Å, c = 5.9697 Å[10]) where the cation sublattice is that of the perovskite structure but the oxygen sublattice can be described as alternating layers of $InO_6$ octahedra and $InO_4$ tetrahedra[9-11]. For higher temperature, BIO undergoes a phase transition from orthorhombic to tetragonal around 925 °C and to cubic around 1040 °C[10]. In the cubic phase the empty oxygen sites are randomly distributed oxygen vacancies, that is mobile charge carriers, which leads to an oxygen ion conductivity by one to two orders of magnitude higher as compared to that of the brownmillerite phase[6, 9]. In relation to this, the empty oxygen sites in the lower temperature phases of BIO (below 1000 °C) are typically referred to as ordered oxygen vacancies, even though they are not actually crystallographic defects. These vacancies in the $InO_4$ tetrahedra plane can be described as ordered in 1-dimensional chains along the [001] direction parallel to the (010) plane of BIO[10-12].

The proton uptake occurs by dissociative water vapour absorption with the OH⁻ ions filling the oxygen vacancies and the proton bonding to an oxygen ion of the brownmillerite crystal structure. The oxygen ion conduction occurs *via* hopping between adjacent oxygen vacancies, and the proton conduction *via* a Grotthuss-type mechanism where the proton is transferred between oxygen ions or rotates around the oxygen ion, reorientating its position between two InO$_6$ octahedra or InO$_4$ tetrahedra[1].

A number of theoretical, X-ray or neutron diffraction, and nuclear magnetic resonance studies focus on the structural and charge transport properties of BIO, and on the establishment of an ordered sublattice of both charge carriers, oxygen ions[12-17] or protons[18-22]. The presence of an ordered charge carrier sublattice would in principle impose severe restrictions to the charge transport and/or induce a significant anisotropic behaviour of the conductivity of BIO. The latter topic has never been investigated experimentally so far[3].

Anisotropic oxygen ion conduction occurs in layered perovskite-structured materials, such as mixed ionic electronic conducting Ruddlesden−Popper phases and brownmillerites[3, 23], and has also been observed in purely oxygen ion conducting mellite or apatite type single crystals[24-28] and the cubic perovskite SrTiO$_3$[29]. In lanthanum polyphosphate[30], in hydroxyapatite[31], and in langatate[32] anisotropic proton conduction was found. The first two materials show a tetrahedral chain-like structure, the latter contains tetrahedral layers and a layered cation sublattice. Thus, these crystal structures are quite different from the hydrated brownmillerite structure of BIO, where layers of cubic perovskite alternate with layers where the octahedral are slightly tilted. Further, for Ga-based oxides with anisotropic oxygen ion conduction, isotropic proton conduction over the Grotthuss-type mechanism in wet atmosphere has been suggested[25].

The hydrated compound of BIO, Ba$_2$In$_2$O$_6$H$_2$, is described by alternating layers of different types of InO$_6$ octahedra[18, 21]. The hydrated material can be dehydrated by a heat treatment under dry conditions. Of specific concern for this work is that the dehydration process upon heating is a two-stage mechanism. The dehydration starts at around 275-300 °C and is characterised by a homogeneous release of protons over the entire oxide lattice up to ca. 370 °C, whereas upon further temperature increase there is a preferential desorption of protons originating in the nominally tetrahedral layers[22]. At around 370 °C, approximately 50% of the protons have left the structure and at 600 °C the material is essentially dehydrated[22]. Such a non-simple dehydration behaviour is consistent with different types of proton sites, which were also predicted theoretically in relation to an order among protons in BIO[18]. Like the order among oxygen ions, an order among protons suggests potentially different proton transport properties along different crystallographic directions in the BIO matrix[22].

In response to the significant number of studies that calculate or measure planar ordering of oxygen ions and (more recently) protons, the ionic conduction in BIO is characterised here along different crystallographic directions. For this purpose, epitaxial thin films of BIO are grown on differently oriented single crystalline substrates and the ionic conductivity is measured in-plane. Different gaseous atmospheres result in either predominant oxygen ion of proton conduction, enabling the investigation of both types of conduction in a single sample. Any anisotropic behaviour is expected to be different for the two charge carrier types since the mechanism of charge transport differs fundamentally. An anisotropic ion conductivity may allow the systematic tuning of the conduction properties in a fashion analogous to that attainable via epitaxial strain or through changes of crystallinity and/or morphology[32-34].

This study reports the first investigation of the anisotropic oxygen ion and proton conductivities of epitaxial BIO films. The oxygen-ion conductivity is investigated also in the temperature region across the order-disorder phase transition. For this purpose the in-plane conductivity of epitaxial BIO thin films of different crystallographic orientations is probed in dry and wet atmospheres in the temperature range of 500-700°C and 200-320°C, respectively.

**Experimental**

The epitaxial thin films were grown by pulsed laser deposition (PLD) on single crystalline MgO substrates. Differently oriented MgO substrates were employed: $10 \times 5 \times 0.5$ mm$^3$ (001)MgO, $10 \times 5 \times 0.5$ mm$^3$ (011)MgO and $5 \times 5 \times 0.5$ mm$^3$ (111)MgO. The disc-shaped BIO pellet, used as target for PLD, was fabricated by solid state synthesis, using starting reactants of $BaCO_3$ (99.999%) and $In_2O_3$ (99.99%), with two heat treatments at 1300 °C for 24 h, followed by spark plasma sintering at 1050 °C, at 75 MPA, for 5 min. The target was ablated in a custom made PLD system with a 248 nm KrF excimer laser (Coherent Lambda Physics GmbH) with a pulse length of 15 ns. A spot size of the laser on the target of 1.9 mm$^2$, a fluence of 1.9 Jcm$^{-2}$, a frequency of 5 Hz, and a target to substrate distance of 8 cm were used. The system reaches a base pressure of $10^{-8}$ mbar and the ablation was carried out in an $O_2$ atmosphere of $10^{-3}$ mbar. A deposition temperature of 710 °C (measured with a pyrometer at the substrate surface) was applied. With the selected deposition parameters the deposition rate of BIO was found to be 0.10 nm s$^{-1}$, as determined by X-ray reflectometry. The thin films studied here were 45-60 nm thick.

The films were analysed with X-ray diffraction with ω/2θ scans showing the crystallographic out of plane orientation and X-ray reflectometry yielding the film thickness using a D500 Siemens diffractometer. Reciprocal space maps were recorded in a Seifert diffractometer equipped with a 1D detector. The chemical composition was probed with Rutherford backscattering using a 4 MeV He beam.

The conductivity of orthorhombic BIO is investigated here along different crystallographic directions in dry Ar, dry $O_2$ and in wet atmospheres using $H_2O/Ar$ and $D_2O/Ar$. Comparing the proton to the deuteron conductivity, it can be determined whether protons are the main charge carriers[33-34]. For the two-point in-plane electrical characterisation, Pt electrodes with a Ti adhesion layer are deposited by magnetron sputtering at room temperature on top of the films. The electrode shape is defined by shadow masks: Rectangular, parallel electrodes form a conduction channel of 1 mm length and 4 mm width. Pt was sputtered with 40 W at $3 \cdot 10^{-2}$ mbar Ar for 2 min (100-200 nm) on a Ti adhesion layer sputtered with 20 W at $7 \cdot 10^{-2}$ mbar for 1.5 min (~5 nm). The target to substrate distance was 4 cm.

The conductivity was measured by impedance spectroscopy with an in-house made set-up in a tube furnace under a gas flow (Ar, $O_2$, $H_2O$/Ar, $D_2O$/Ar). The electrodes are wired to the read-out electronics using Ag or Pt paste and Au wires. A Solartron 1260 Impedance gain phase analyser was used with a bias voltage of 1 V in the frequency range between 1 Hz and 1 MHz with integration times of 1-5 s. For measuring the proton or deuteron conductivity, the samples were first dehydrated in dry Ar atmosphere at a temperature of 600 °C, then cooled to 400 °C. At this temperature, the gas was switched to Ar bubbled through deionised water or $D_2O$ (99.9%, Merck KGaA) and the temperature was decreased with 100°C/h to hydrate the samples. To maximize the degree of hydration, the samples were kept overnight under a flow of wet Ar at 250 °C. For the conductivity measurements the temperature was not increased above the growth temperature of 710 °C so that the film remained unchanged. Increasing the temperature further might result in a range of, for example, morphological or compositional changes. The lower temperature limit was determined by the maximal resistance measurable with the set-up (~500 MΩ). For one sample the temperature range for the conductivity measurements in Ar atmosphere was extended up to 1000 °C after all other measurements were concluded to probe the conductivity in the temperature range around the orthorhombic to cubic phase transition of BIO.

**Results and discussion**

**Structure and composition.** MgO single crystals are used as substrates for the growth of epitaxial BIO thin films. The cubic perovskite phase of BIO (a=4.274 Å[10]) exhibits a good lattice matching for epitaxy with the rock-salt structure of MgO ($a_{MgO}$=4.212 Å). Moreover, MgO is a highly insulating material (electronic and ionic) and therefore well suited as a substrate for measuring the ion conductivity in thin films at high temperatures in-plane, i.e. along the direction of the substrate surface. Also the orthorhombic perovskite phase of BIO exhibits good lattice matching with MgO assuming a cube-on-cube growth of the pseudocubic unit cell of the orthorhombic BIO phase on the cubic unit cell of MgO. In Fig. 1 the expected epitaxial orientations and the crystal structure are sketched for the growth of BIO on (001), (011) and (111) out-of-plane oriented MgO substrates.

The surface lattice of a (001)MgO substrate consists of squares and the double of $a_{MgO}$ fits to the (10-1) diagonal of the orthorhombic unit cell of BIO, so that BIO should grow (101) out-of-plane oriented on (001)MgO (Fig. 1a and 1e). Domains will form which are 90° rotated with respect to each other because the surface lattice of (001) MgO has a 90° rotation symmetry.

The surface lattice of (011)MgO is rectangular, and the diagonal of the MgO unit cell fits best to the lattice parameter c of BIO, while four times $a_{MgO}$ matches b. As a consequence, an epitaxial film of BIO on (011)MgO is expected to be (100) out-of-plane oriented (Fig. 1b and 1f). The surface lattice of (011)MgO only shows a 180° rotation symmetry, which would allows selecting the direction of the conductivity measurement to be either parallel (along the (01-1)MgO in-plane direction) or perpendicular (along the (100)MgO in-plane direction) to the $InO_4$ planes in BIO. Either ways the orientation of the vacancy channels would be perpendicular to the direction of conduction (Fig. 1f).

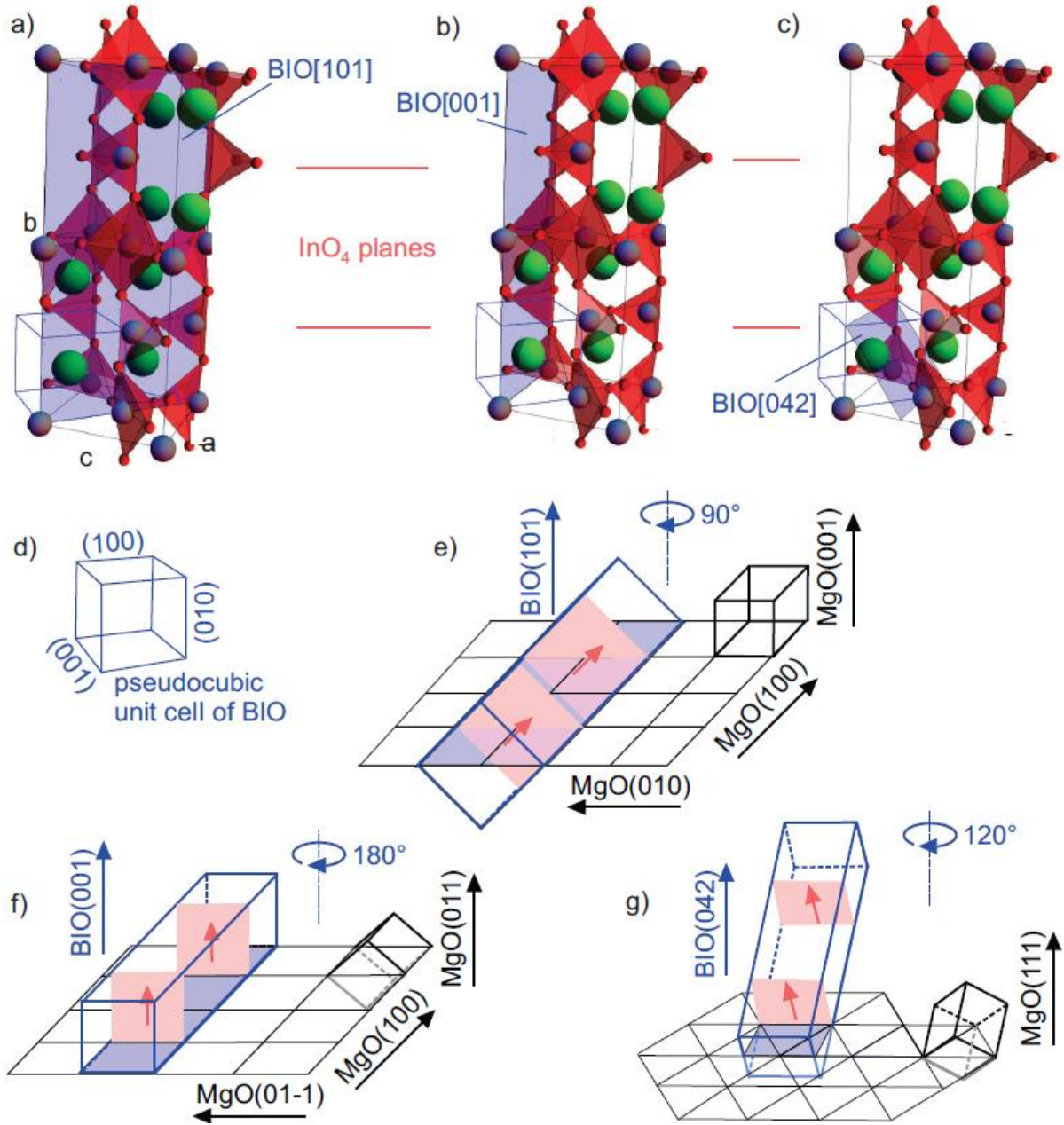

***Figure 1.*** *The expected epitaxial orientations of BIO on (001), (011) and (111) MgO substrates are sketched. a-c) show the atomistic structure of the orthorhombic BIO unit cell (Ba green, In grey and O red), and the pseudocubic unit cell is indicated (sketched separately also in d). The planes in the orthorhombic unit cell [101], [100] and [240] are parallel to the [001], [011] and [111] planes of the pseudocubic unit cell. The positions of the InO$_4$ tetrahedra planes are indicated with pink lines. The black grids in e-g) show the lattice of the MgO surface, i.e. the cut of the MgO unit cells parallel to the substrate surface. The complete unit cell of MgO is shown on the right side of the grids. Assuming the cube-on-cube growth of the pseudocubic BIO unit cell on MgO, the orientation of the BIO unit cell with respect to the substrate surface is sketched in e-g) in blue. The InO$_4$ tetrahedra planes (the oxygen vacancy planes) are indicated in pink, the arrows show the orientation of the 1-dimensional vacancy channels. Due to the rotation symmetry within the substrate surface lattice (black), the epitaxial order is preserved when the BIO unit cell is rotated with respect to the MgO substrate normal by e) 90°, f) 180° and g) 120°.*

The surface lattice of (111)MgO is hexagonal. Again assuming the cube-on-cube matching of the pseudocubic BIO unit cell with the MgO unit cell, the BIO film on (111)MgO is expected to exhibit the (240) out-of-plane orientation with three types of domains, corresponding to the 120° rotation symmetry of the hexagonal surface lattice of (111)MgO, as sketched in Fig. 1c and 1g.

In the temperature range of the orthorhombic phase, epitaxial BIO films were successfully deposited by pulsed laser deposition on MgO substrates with (001), (011) and (111) out-of-plane orientation as shown by the X-ray diffraction analysis of the thin films in Fig. 2.

Rutherford backscattering yielded a Ba:In cation ratio of 0.95 with an error of 1%, showing a slight Ba deficiency.

The ω/2θ scans (Fig. 2a) of the BIO films only show the reflections that were expected for the epitaxial growth on the respective substrates (Fig. 1), though there is indication for twinning on the MgO(011) substrate. The epitaxial orientation is further confirmed by reciprocal space mapping where in all cases assymetric reflections were found in the expected angular range.

On the (001)MgO substrate, the (202) reflection of orthorhombic BIO is visible as a shoulder of the substrate peak (Fig. 2b) while the (101) and (303) reflections are not visible due to their small relative intensity[10]. The reciprocal space map of the assymetric (343) and (363) reflections of BIO is shown in Fig. S1a in the supporting information.

On the (011)MgO substrate, the (200)/(002) BIO reflections are clearly visible while the (400)/(004) reflections coincide with the substrate reflection. Fig. 2c shows higher resolution measurements of relevant film reflections. Due to the similarity of the lattice parameters a and c of BIO the twining is clearly visible, though the (200) reflection seems to be shifted to lower 2θ values. While c is very close to the diagonal of the MgO cube, a is larger, so that this shift corresponds to the resulting compressive strain. To confirm that there is indeed no 90° rotation symmetry in the BIO films on (011) oriented MgO substrates, the asymmetric (240)/(042) reflection of BIO is mapped with the X-rays parallel to MgO(0-11) and, rotating the substrate by 90°, with the X-rays parallel to MgO(100). The latter configuration showed no diffraction peak, as shown in Fig. 2d.

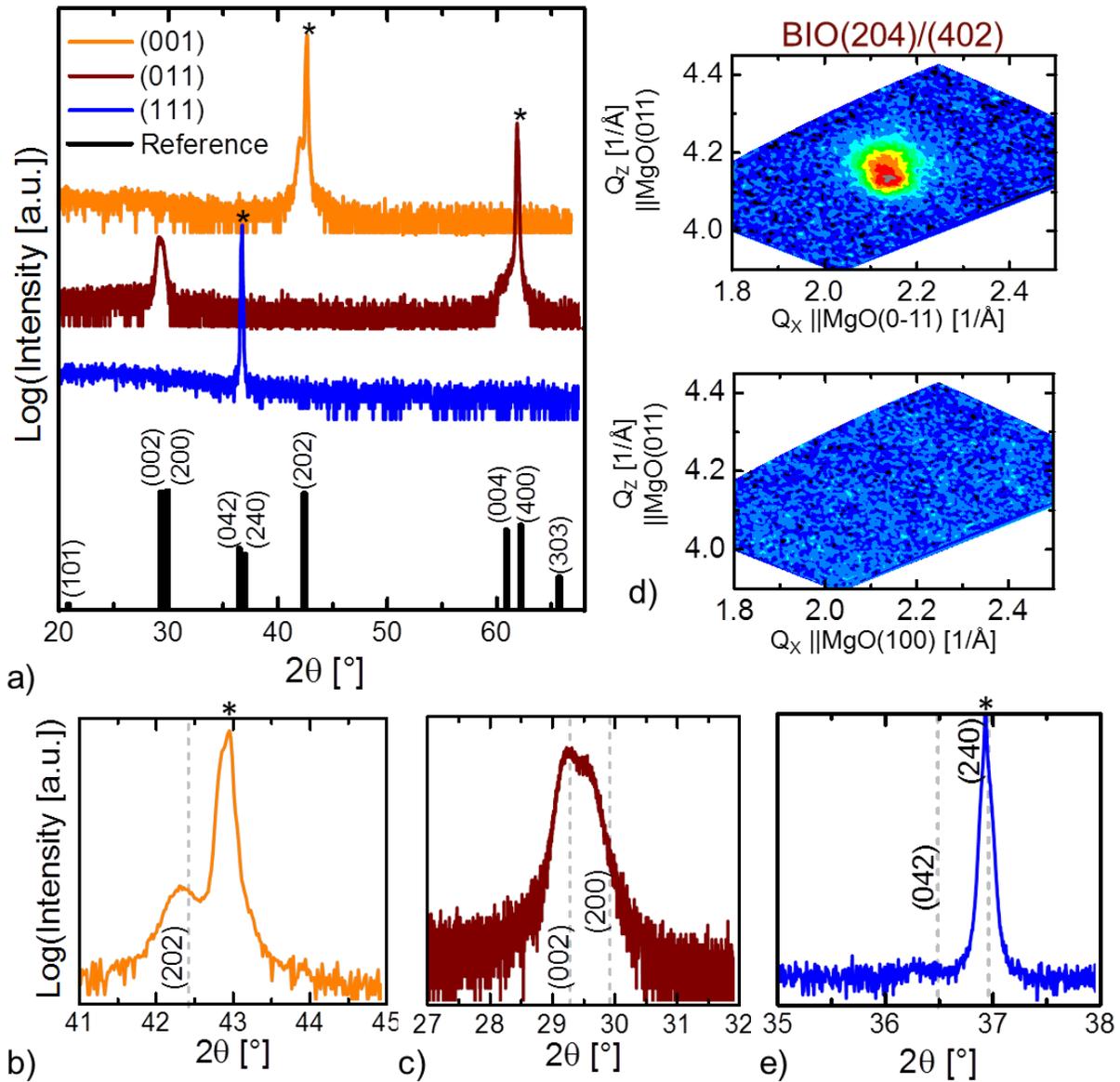

*Figure 2.* XRD analysis of epitaxial BIO films: (a) ω/2θ scans of epitaxial BIO films grown on (001), (011), and (111) oriented MgO substrates, (b-e) high-resolution XRD measurements for selected angular ranges of the three samples. The (202)BIO reflection on MgO (001) (b) and the (002)/(200)BIO reflections on MgO(011) (c) are clearly visible. (d) shows reciprocal space maps of the BIO(240)/(042) reflection in the BIO film on MgO(011), once aligned for the expected epitaxial order, i.e. X-rays||MgO(0-11) and once, 90° rotated, with the X-rays||MgO(100). (e)

*shows the high resolution measurement around the (042)/(240)BIO reflection on MgO(111). Substrate peaks are labelled in with asterisks and reflexion positions based on a powder diffraction pattern[10] are shown as a reference. Additional reciprocal space maps of assymetric reflections on MgO(001) and MgO(111) are shown in the supporting information Fig. S1.*

On (111)MgO, the (042) reflection of the BIO film is very close to the substrate reflection. The BIO(042) reflection has a 50 times smaller relative intensity than for example the BIO(200)/(002) reflections[10] and would therefore be close to the noise level. Even using higher resolution and longer integration times (Fig. 2e), the (042) reflection is barely distinguishable from the noise. As twinning occurs on the MgO(011) substrate, it may also occur here. However, the (240) reflection is not stronger than (042) and coincides directly with the substrate reflection. In the supporting information, Fig. S1b shows the reciprocal space map of the assymetric (033)(330) reflection, confirming the expected epitaxial relationship.

In summary, though the twinning precludes the characterisation along the 1D vacancy channels, epitaxial BIO thin films were fabricated so that the conduction can be measured along different crystallographic directions, e.g. parallel or perpendicular to the oxygen vacancy planes.

**Oxygen ion and proton conductivity of epitaxial BIO films.** Based on the orientation of the electrodes (Fig. 3) we have measured the conductivity $\sigma_{ion}$ along different crystallographic directions in BIO, namely along the (121), the (100)/(001), the (010) and the (1-20)/(0-21) direction or equivalent, taking into account the rotation symmetry (Fig. 1). The (010) direction crosses the oxygen vacancy planes of the InO$_4$ tetrahedra (pink in Fig. 1) at 90°.

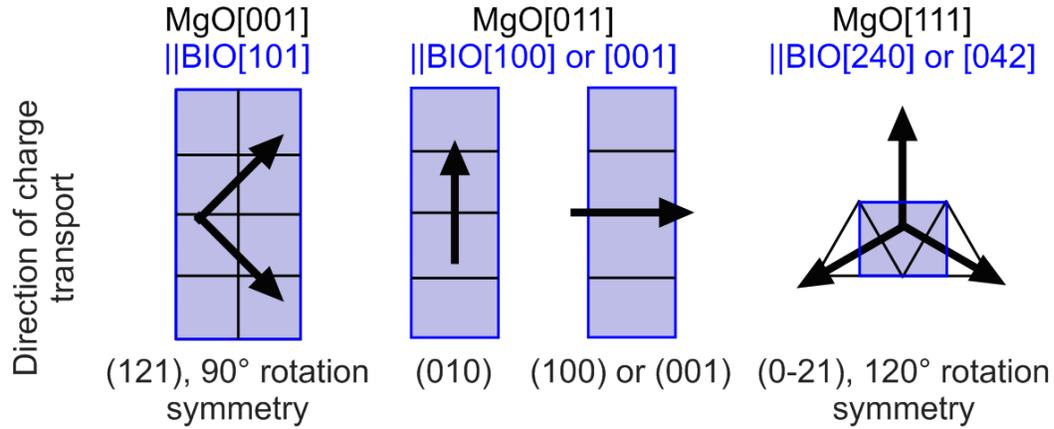

*Figure 3. The direction of charge transport lies in the planes parallel to the MgO substrate surface (black grid), that is, in the BIO[101], [100]/[001] or in the [240]/[042] planes. The blue area shows the cut of the unit cell of BIO, which is parallel to the substrate surface, as seen in Fig. 1. The arrows denote the direction of the charge carriers based on the position of the parallel patch electrodes.*

The (100) and (001) directions are parallel to these planes, but due to the twinning, the vacancy channels will in some domains be parallel and in others perpendicular to the direction of conduction. The (121) and (0-21) directions cross the InO$_4$ tetrahedra planes at angles of 45° and around 75°, respectively. On the MgO(001) substrate, the electrodes are placed such that BIO(121) is selected as the direction of conduction. That way the angle between the tetrahedral planes and the direction of conduction is the same in both types of domains (which are 90° rotated with respect to each other).

Fig. 4 reports the conductivity measurement in different gaseous environments. Conductivity data reported in literature for dry and wet atmospheres are shown in Fig. 4(a). The measurements were conducted on polycrystalline sintered ceramic pellets and mostly the bulk conductivity is reported[7-9]. The proton conductivity in wet atmosphere below 300 °C is about two orders of magnitude higher than in dry Ar and dry $N_2$ where BIO shows predominant oxygen-ion conductivity. At higher temperatures (~600 °C), the curve for wet atmospheres merges with those measured in dry environments due to the dehydration of the sample. The large change in conductivity at the order-disorder transition of the oxygen vacancy sub-lattice is clearly visible above 900 °C. It is noted that the proton conductivity of BIO has also been reported in $H_2$-containing gas, in which case proton uptake introduces additional electrons. Using nanocrystalline samples (40 nm grain size), BIO was found to be chemically unstable in $H_2$ containing atmospheres above 450 °C[5]. Further, the conductivity was 4 orders of magnitude higher[5] than reported in water vapour[6], which was partially attributed to grain size effects[35].

For the oxygen ion conduction measured in dry Ar, the epitaxial BIO thin films show a conduction behaviour very close to other literature reports[1,3] of the bulk conductivity (Fig. 4(b)). For the proton conduction measured in wet Ar, the conductivity of the epitaxial BIO films lies about two orders of magnitude above the data reported in literature for the sintered ceramic pellets measured in $H_2O$/air[6, 8] (Fig. 4(c)). For comparison, the conductivity measured for instance along the (121) crystallographic direction of BIO at 300 °C is a factor of ~6 smaller than the bulk conductivity of the well-known proton conducting SOFC electrolyte 20% Y-doped $BaZrO_3$ at the same temperature[36].

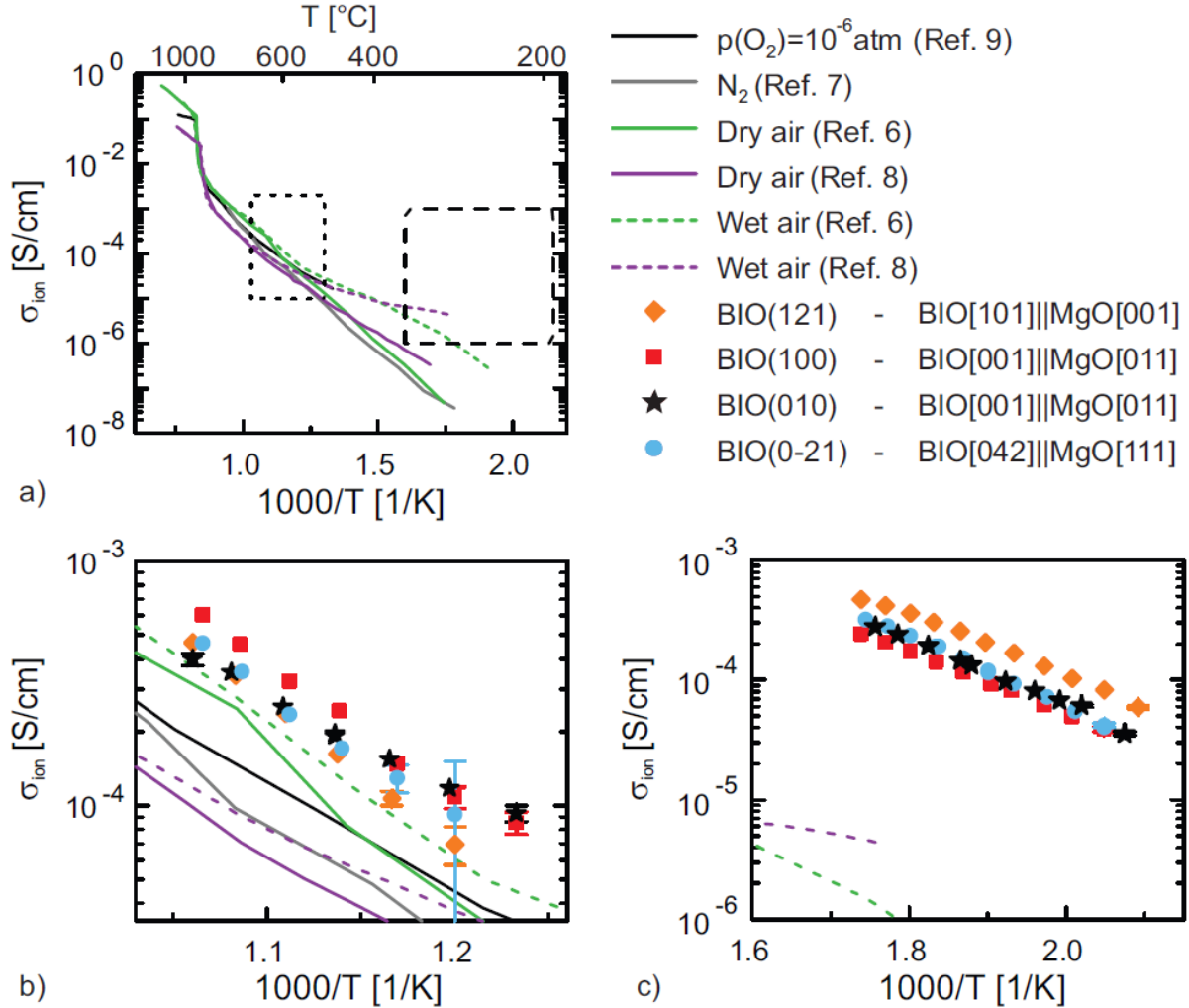

*Figure 4.* Electrical characterisations in different atmospheres are compared to literature data[6-9] for the conductivity of polycrystalline BIO in dry (solid lines) and water vapour containing atmospheres (dashed lines) in the temperature range 1100-200 °C (a). The temperature ranges investigated for this study in dry and wet atmospheres are indicated by the areas given by dotted and dashed lines, respectively. The conductivity of the BIO epitaxial films along different crystallographic directions in dry Ar (b) and in wet Ar ($H_2O$/Ar) (c) are shown. The direction (hkl) of conduction and the planes [hkl] of BIO and MgO along which the conductivity is measured are indicated. Error bars from the fit of the impedance spectra (Fig. S4) are indicated for the points with low conductivity as long as the error bar is larger than the respective symbol.

The green dashed line in Fig. 4c reports the total conductivity of BIO polycrystalline samples[6]. We may thus assume that the much lower conductivity arises from poorly conducting grain boundary regions, as it is the case for other proton conducting oxides[37].

The conductivity of the BIO pellets also depends on sample history[6]. Using polycrystalline pellets, the highest proton conductivity was obtained when the sample was hydrated for 1000 h at 250 °C and the conductivity was measured increasing the temperature stepwise[6]. This is similar to our hydration process, except that thin films need much shorter equilibration times. In contrast, hydrating the sample for 48 h at 600 °C and measuring the conductivity during cooling led to the lowest proton conductivity[6]. This may explain the low bulk proton conductivity and the much smaller activation energy measured using sintered ceramic pellets upon cooling down the sample from a maximum temperature of 1000 °C[8] (purple dashed line in Fig. 4c). In this last case, proton uptake may still be ongoing during cooling.

The large difference among the reported results certainly demands further investigation of the grain interior conductivity of BIO. However, we suggest that the conductivity measured using highly ordered epitaxial thin films may be closer to the intrinsic grain interior conductivity of fully hydrated BIO. This statement is supported by the high grain interior proton conduction measured for epitaxial thin films of Y-doped $BaZrO_3$[37] in the same temperature range used in the present study.

Above 300 °C the conductivity no longer increased due to dehydration. Decreasing again the temperature, the previous conductivity values were recovered with time. The onset of the dehydration in the epitaxial thin films is similar to that reported for polycrystalline pellets[6, 8].

In dry $O_2$ (see Fig. S2), the conductivities of the BIO films are higher than in dry Ar which is in agreement with the influence of the oxygen partial pressure on the BIO conductivity in dry environments. It has been argued that the incorporation of excess oxygen results in p-type electronic conduction[9, 11]. In typical oxygen ion conductors like doped $CeO_2$ or Y-stabilised zirconia, the conductivity is constant down to oxygen partial pressures of ~$10^{-7}$ mbar around 600-800 °C[38-39]. In the same temperature range, the conductivity of typical proton conductors like Y-doped $BaZrO_3$ significantly increases for oxygen partial pressures above ~$10^{-2}$ mbar (oxygen partial pressure in dry Ar)[40-41]. This is due to the progressively increased p-type electronic contribution arising from the incorporation of oxygen ions into the vacancies[41]. In this respect, BIO resembles more closely a typical proton conductor than a typical oxygen ion conductor.

Comparing the conductivity in $H_2O$/Ar and $D_2O$/Ar, the isotope effect is observed for all three out-of-plane orientations (Fig. S3). Within the experimental uncertainty, the effect is the same for the three samples, as should be expected. A difference in activation energy of ~0.03(1) eV and a pre-exponential factor ratio $\sigma_0^{H^+}/\sigma_0^{D^+}$ of ~1.00(3) was measured. These values are in the range reported for the bulk properties of other proton conducting oxides[33-34]. The fact that $\sigma_0$ is basically unchanged shows that the vibrations of the oxygen sublattice determine the attempt frequency for proton transport rather than the vibrations of the proton-oxygen bond, as expected for typical perovskite proton conductors[42]. The observation of the isotope effect confirms, as expected from previous reports, that protons are the dominant charge carriers in a humidified environment.

The order-disorder phase transition in BIO is known to occur around 900-1000 °C[6, 8-10]. This is above the deposition temperature of 700 °C, therefore the BIO film may undergo structural changes apart from the structural transition. The films may for example lose the epitaxial ordering in this temperature range. To complete this investigation, one sample (conduction along (0-21)) was measured in dry Ar atmosphere up to 1100 °C (Fig. 5) after all other measurements in dry and wet atmospheres were completed. Below the deposition temperature of 700 °C, the conductivity is in line with the first electrical characterisation in the same gaseous environment (light blue dots in Fig. 5). This also shows that the measurements in $O_2$, $H_2O$/Ar and $D_2O$/Ar did not alter the sample. Raising the temperature above 700 °C, we did not observe a sharp transition, as reported in the literature for pellets, but rather a gradual increase of the slope of the conductivity curve. The slope decreased again reaching a constant value above about 960 °C.

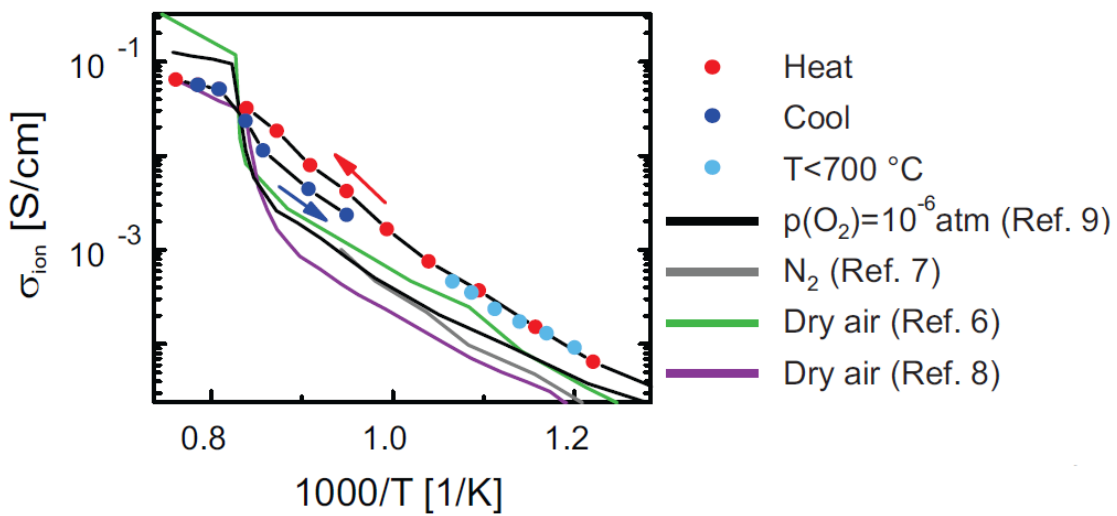

*Figure 5. Conductivity measurement of the BIO film on MgO(111) where the phase transition was probed in dry Ar. The conductivity recorded during heating and cooling is indicated in red and blue respectively. The previous measurement with a maximal temperature of 700 °C (Fig. 3) is shown in comparison with literature data[6-9].*

Upon cooling, the conductivity reaches again the same value at ~960 °C as during heating, after that it drops and then approaches the low temperature data, exhibiting a similar slope as the initial measurement (blue in Fig. 4(b)). During cooling, there is no reason for the film to change anymore, so this drop in the conductivity corresponds to the order-disorder transition. It appears that for thin films the transition is not as sharp as that reported in literature for sintered ceramic pellets[6, 8-9]. This different behaviour originates probably from epitaxial constraints through the substrate or less volumetric constraints out-of-plane.

**Anisotropic ion conductivity.** As shown in Fig. 4 the oxygen ion and the proton transport depend on the crystallographic plane along which the conductivity is measured. As an example, the oxygen ion conductivity at 560 °C varies within a factor of ~1.7 among the different crystallographic directions investigated in this study. This effect is far smaller than the effect of anisotropy reported for example for $Pr_2NiO_{4+\delta}$ or $Nd_2NiO_{4+\delta}$[23]. The structure of these materials consists of alternating perovskite and rock-salt layers and the conductivity across the layers is 3-2 orders of magnitude lower than that along the layers at around 600 °C[23].

For BIO, the activation energy $E_A$ and the pre-exponential factor $\sigma_0$, which determine the conductivity according to the Arrhenius equation, are compared in Fig. 6 along different crystallographic directions for the oxygen ion and proton conductivities. The data shows that a significant anisotropy can be observed also for the oxygen ion conductivity in BIO. By comparing the conductivity in dry Ar along the (121) and (010) directions of BIO, $\sigma_0$ decreases by about a factor of 500 and the activation energy by about 0.5 eV.

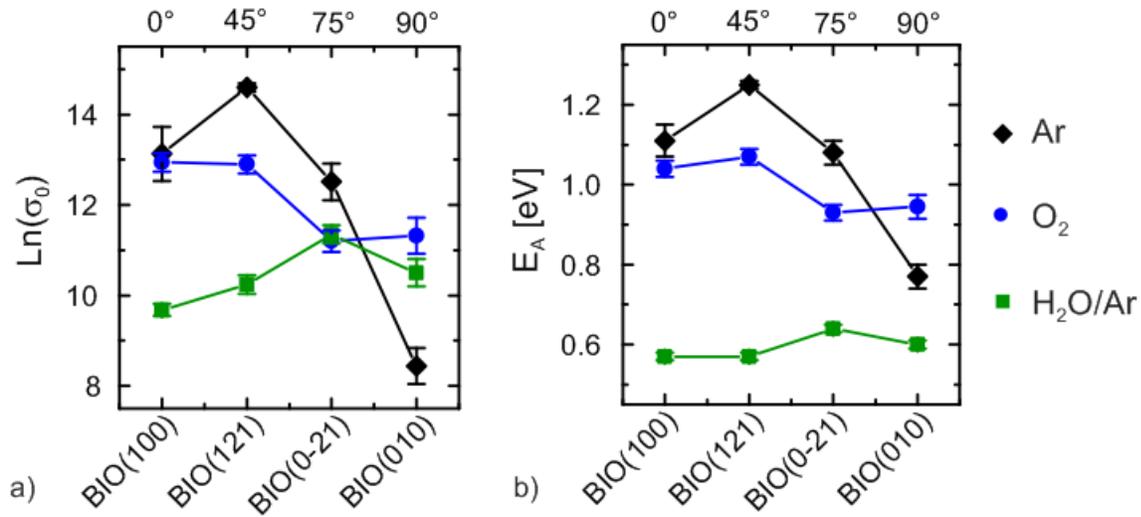

*Figure 6. Pre-exponential factor (a) and activation energy (b) along different crystallographic directions in different gaseous environments. The angle between the direction of conduction and the oxygen vacancy planes is indicated on the top x-axis.*

As discussed above, the brownmillerite structure of BIO can be visualized as a layered structure of alternating layers of $InO_6$ octahedra, where all oxygen ion sites are occupied, and of $InO_4$ tetrahedra where the lattice sites which would complete the tetrahedra to octahedra can be considered as oxygen vacancies. The data in Fig. 6 is ordered with respect to the angle between the direction of conduction and the oxygen vacancy planes (see Fig. 1) along the $InO_4$ layers.

Fig. 6 shows that both $E_A$ and $\sigma_0$ have a minimum for conductivity measurements along the (010) direction of BIO which is perpendicular to the $InO_6$ – $InO_4$ layers. This indicates that even though the potential well to be overcome for hopping along this crystallographic direction is relatively small (smaller $E_A$), the probability of hopping is also small (small value of $\sigma_0$ which is proportional to the number of free sites an ion can hop into). We conclude that it is unlikely for an oxygen ion to hop into the $InO_6$ octahedra layers since, due to the absence of oxygen vacancies, this would correspond to a correlated jump of another oxygen ion that simultaneously leaves the layer.

The 1-dimensional vacancy channels along the [101] planes of BIO discussed above are expected to result in a highly anisotropic behavior of the oxygen ion conductivity since such vacancy channels should provide preferential conduction pathways[3]. For instance $\sigma_0$ along BIO(100) was expected to be relatively small since this direction of conduction is along the vacancy planes of the $InO_4$ tetrahedra (as shown in Fig. 1f and indicated in Fig. 6) but perpendicular to the direction of the vacancy channels. Here, however, the twinning on MgO(011) and the domain formation due to the substrate symmetry on MgO(001) unfortunately precludes the direct characterization of the conductivity along the vacancy channels.

The (101) out-of-plane oriented BIO film is the case where the 1-dimensional vacancy channels have the largest vector component parallel to the substrate surface. The angle between the vacancy channels and the (121) in-plane direction of conduction is 45° as reported in Fig. 6 in both possible domains of BIO, which are rotated by 90° with respect to each other (see Fig. 1 and 3). Assuming an overall zigzag migration path following the vacancy channels through adjacent domains, this is most likely the reason why $\sigma_0$ is the largest for the conduction along the (121) direction.

For the oxygen ion conduction, the activation energy changes by 0.48 eV comparing BIO(121) and BIO(010). A similar change of activation energy along different crystallographic directions was observed for $Pr_2NiO_{4+\delta}$ while the activation energy in $Nd_2NiO_{4+\delta}$ was unchanged within the error bar of the measurement[23]. In other examples of anisotropic oxygen ion conduction, the activation energy differs by 0-0.14 eV[24, 26].

Since $\sigma_0$ is proportional to the number of free sites, it was possible to discuss the anisotropy of $\sigma_0$ based on the order of oxygen ions. But to gain more insight in the variation in $E_A$ between different crystallographic directions, theoretical simulations should be employed in the future.

The proton conduction is also anisotropic, though to a lesser degree than the oxygen ion conduction. In wet Ar atmospheres it is observed that $\sigma_0$ varies by up to a factor of 5 while $E_A$ lays always around 0.6 eV and changes by 0.07 +/- 0.02 eV. This means that the energy barriers for proton transport by the Grotthuss mechanism from one oxygen ion to another does not significantly vary along the different crystallographic directions (Fig. 6). The structure of fully hydrated BIO contains alternating layers of untilted and slightly tilted oxygen octahedra. It appears that this slight deviation from a cubic perovskite does not induce an anisotropy of the energy barriers. It does, however induce an order among protons according to theoretical predicitons, as discussed in literature[21-22]. Like the order among oxygen ions, this is expected restrict the number of free sites available for jumps in a certain direction, resulting in the anisotropy of $\sigma_0$ (Fig. 6).

A similarly anisotropic behaviour as in BIO was found in other experimental studies on anisotropic proton conduction. In langanate, the activation energy is unchanged while the conductivity varies by a factor of 1.5[32]. The non-hydrated langanate also has an oxygen sublattice of tetrahedral and octahedral layers like the brownmillerite, but the cation lattice is layered. In polyphosphate materials, which have a structure resembling chains of tetrahedra, the conductivity also varied by a factor of 5, while the activation energy changed by 0.05 eV[30].

In these two examples the anisotropic behaviour of the proton conductivity arises from the presence of preferential conduction pathways offered by the host lattice. In BIO however, the anisotropic proton conduction most likely arises from the theoretically predicted proton ordering.

**Conclusions**

The anisotropic oxygen ion and proton conduction in $Ba_2In_2O_5$ is characterized here by using epitaxial thin films and thereby also the first investigation of the conductive properties of epitaxial BIO thin films is presented. We find that the pre-exponential factor is strongly anisotropic for the oxygen ion conduction. Reasons for the anisotropic oxygen ion conduction are suggested in relation to the order of the oxygen vacancies in the brownmillerite structure. Combined with the anisotropy of the activation energy, however, the conductivity in the temperature range under investigation does not strongly vary along different crystallographic directions. Due to twinning of (100)/(001)BIO grown on (011)MgO resulting from the similarity of the lattice parameters a and c of orthorhombic BIO, the conductivity cannot be probed selectively along the 1-dimensional oxygen vacancy channels. We do not see at present how to overcome this problem using epitaxial thin films, but it would be possible by growing a single crystal of BIO.

The proton conduction is anisotropic as well which is attributed to an ordering among protons reported in earlier studies. Nevertheless, a similar degree of anisotropy of the proton conduction was observed along different crystallographic directions in materials where proton ordering is not present but where protons find preferential conduction pathways through chain-like or layered structures.

SUPPORTING INFORMATION

In the supporting information, additional reciprocal space maps are shown as well as additional data of the electrical characterization. The latter includes the conductivity measurements in dry oxygen, the isotope effect and examples for complex impedance plane plots.

ACKNOWLEDGMENT

The research leading to these results has received funding from the Swiss National Foundation for Science (SNFS) under grant agreement No. 200021_153362, and the Swedish Research Council under grant agreements No. 2010-3519 and 2011-4887. We thank Max Döbeli of the Laboratory for Ion Beam Physics at ETH Zurich (CH) for the compositional analysis by Rutherford backscattering and Ulrich Häussermann at Stockholm University (SE) for the spark plasma sintering of the BIO sample.